\documentclass[conference]{IEEEtran}

\usepackage{amsmath}
\usepackage{cite}
\usepackage{amsmath,amssymb,amsfonts}
\usepackage{algorithmic}
\usepackage{graphicx}
\usepackage{textcomp}
\usepackage{xcolor}

\usepackage{algorithm}
\usepackage{algorithmic}

\DeclareMathOperator*{\argmin}{argmin}

\usepackage{hyperref}

\usepackage{amsthm}
\newtheorem{theorem}{Theorem}[section]

\begin{document}

\title{On Study of Mutual Information and Its Estimation Methods}

\author{\IEEEauthorblockN{Marshal Arijona S.}
\IEEEauthorblockA{Faculty of Computer Science \\
University of Indonesia\\
Depok, Indonesia\\
Email: marshal.arijona01@ui.ac.id}}

\maketitle

\begin{abstract}
The presence of mutual information in the research of deep learning has grown significantly. It has been proven that mutual information can be a good objective function to build a robust deep learning model. Most of the researches utilize estimation methods to approximate the true mutual information. This technical report delivers an extensive study about definitions as well as properties of mutual information. This article then delivers some reviews and current drawbacks of mutual information estimation methods afterward.

\end{abstract}

\IEEEpeerreviewmaketitle

\section{Introduction}
Mutual information (MI) is viewed as one of the most fundamental measurements to quantify the dependence of two random variables \cite{club}. Evidently, mutual information has been applied in wide spectrums, including statistics \cite{club, mi-ex-stat-1, mi-ex-stat-2}, biostatistics \cite{club, mi-ex-bio-1, mi-ex-bio-2}, robotics \cite{club, mi-ex-robo-1, mi-ex-robo-2}, and machine learning \cite{infogan, dim}. This shows that mutual information can capture the notion of dependence on nature universally.

For machine learning applications (especially deep learning), MI is used as an objective function or a regularizer in loss function \cite{club}. The objective function is either maximizing the MI or minimizing the MI. MI maximization is applied in various tasks, including representation learning\cite{dim, discrete-representation, club}, generative models \cite{infogan, club}, and reinforcement learning \cite{mi-reinforcement, club}. Meanwhile, MI minimization has taken parts in disentangled representation learning, style transfer \cite{style-transfer, club}, and information bottleneck\cite{information-bottleneck, club}. 

Almost all MI maximization or MI minimization do not use the exact MI but rather compute the estimation. This due to the required closed form of the density function and tractable log-density ratio between the joint distribution and the product of marginal distribution \cite{club}. In the real world, it is not always possible to have all access to the required distributions. Commonly, we only have samples from the joint distribution \cite{club}. Therefore, the estimation methods are proposed to solve the problems. Info-GAN for example is using Barber-Agakov lower bound \cite{barber-agakov} to estimate the mutual information between the latent factor and the generated images \cite{infogan}. Another example is the contrastive predictive model, which uses noise contrastive estimation to estimate mutual information between the current context and the data at the time steps ahead \cite{cpc}. Mutual information estimation is currently active research in machine learning and still opens a huge possibility to improve.  

This article aims to deliver a theoretical study about mutual information. Especially, the article focus on discussing MI from an information theory perspective. Aside from that, the article also reviews some MI estimation methods. The article is represented as follows. In the beginning, the article discusses the background of this article. The preliminaries section helps the reader to understand the basic concepts of information theory. The MI: definitions and properties section is divided into several subsections. The first subsection talks about the definition of mutual information in general. The rest of the subsections talk about the properties of MI, including the convexity and continuity of MI, the consequences of Jensen inequality for MI, the relations between MI and conditional independence distribution, geometric interpretation of MI, and variational form of MI. The MI: estimation methods section delivers a review of several mutual information methods and their current drawbacks. 

\section{Preliminaries}
Sufficient knowledge about entropy and divergence is needed to have a better understanding of mutual information.

\subsection{Entropy}
Entropy can be viewed as a tool to measure the uncertainty of random variable (RV) \cite{cover-2006}. Let $X$ be a discrete random variable on space $\mathcal{X}$ with distribution $P_{X}$. Also, let $x \in \mathcal{X}$ be an element from space $\mathcal{X}$.The entropy of $X$ can be written as:

\begin{align*}
    H(X) &= - \mathbb{E} \left[ \log P_{X}(x) \right] \\
    &= - \sum_{x \in \mathcal{X}} P_{X}(x) \log P_{X}(x)    
\end{align*} Note that the equation is also hold for continuous random variable. The logarithm term in the equation uses either base 2 (\textit{bit}) or base $e$ (\textit{nat}) \cite{cover-2006}. Furthermore, it is easy to see that $H(X) \geq 0$ is satisfied since $0 \geq P_{X}(x) \geq 1$.

Entropy can also be used to measure the uncertainty for more than 1 random variable. Let $Y$ be another discrete random variable on space $\mathcal{Y}$ with distribution $P_{Y}$. At first, we review joint entropy between random variables $X$ and $Y$. Joint entropy $H(X, Y)$ with a joint distribution $P_{X, Y}$ is defined by:
\begin{align*}
    H(X, Y) &= - \mathbb{E} \left[ \log P_{X, Y}(x, y) \right] \\
    &= - \sum_{x \in \mathcal{X}} \sum_{y \in \mathcal{Y}} P_{X, Y}(x, y) \log P_{X, Y}(x, y)    
\end{align*} Then, we define conditional entropy of $X$ given $Y$ with conditional distribution $P_{X|Y}$ as:
\begin{align*}
    H(X | Y) &= \mathbb{E}_{y \in \mathcal{Y}} \left[ H(P_{X|Y=y}) \right] = - \mathbb{E} \left[ \log P_{X|Y}(x|y) \right ]\\
    &= - \sum_{y \in \mathcal{Y}} P_{Y}(y) \sum_{x \in \mathcal{X}} P_{X|Y}(x|y) \log P_{X|Y}(x|y) \\
    &= - \sum_{y \in \mathcal{Y}} \sum_{x \in \mathcal{X}} P_{X, Y}(x, y) \log P_{X|Y}(x|y)
\end{align*} The conditioning impacts on the reduction on entropy means that $H(X) \geq H(X|Y)$ \cite{cover-2006}. We discuss about this inequality in the later section. 

Joint entropy $H(X, Y)$ can be derived from marginal entropy $H(X)$ and conditional entropy $H(Y|X)$.  
\begin{theorem}\label{entropy-chain-rule}
Both $H(X, Y)$ and $H(X|Y)$ derive chain rule property written as:
\begin{align*}
    H(X, Y) = H(X) + H(Y|X) \leq H(X) + H(Y)
\end{align*} 
\end{theorem}
\noindent Note that the inequality holds from the conditioning of $H(Y|X)$. We also can extend the relations for more than two random variables as we call conditional joint entropy. Let us specify another random variable $Z$ on space $\mathcal{Z}$. We can write conditional joint entropy $H(X, Y | Z)$ as:
\begin{align*}
    H(X, Y | Z) = H(X|Z) + H(Y|X, Z) \leq H(X) + H(Y)
\end{align*} with the inequality $H(Y|X, Z) \leq H(Y)$ holds for the equation.

\subsection{Divergence}
Divergence (also known as Kullback-Leibler (KL) divergence or relative entropy) is a measurement of the distance between two distributions over a random variable \cite{notes}. We already specified random variable $X$ on space $\mathcal{X}$ and distribution $P_{X}$. Then, let $Q_{X}$ be another distribution function quantifying RV $X$. KL-Divergence between $P_{X}$ and $Q_{X}$ is defined by:
\begin{align*}
    D_{KL}(P_{X}||Q_{X}) &= \mathbb{E} \left[ \log \frac{P_X(x)}{Q_X(x)} \right] \\
    &= \sum_{x \in \mathcal{X}} P_{X}(x) \log \frac{P_{X}(x)}{Q_{X}(x)} \: (discrete) \\
    &= \int P_{X}(x) \log \frac{P_X(x)}{Q_X(x)} dx \: (continuous)
\end{align*} There are two constraints for the above definitions:
\begin{itemize}
    \item $0 . \log \frac{0}{0} = 0$
    \item $\exists x : Q_{X}(x), P_{X}(x) > 0 \implies D_{KL}(P_{X} || Q_{X}) = \infty$
\end{itemize}
Note that KL divergence is not symmetric means $D_{KL}(P_{X} || Q_{X}) \neq D_{KL}(Q_{X} || P_{X})$. Furthermore, we can also extend KL-divergence into conditional case where probability function $P_X$ is given. In particular, KL-divergence between $P_{Y|X}$ and $Q_{Y|X}$ (not symmetric) given $P_{X}$ can be written by:
\begin{align*}
    &D_{KL}(P_{Y|X}||Q_{Y|X} | P_X) = \mathbb{E}_{P_X} \left[D(P_{Y|X=x}||Q_{Y|X=x}) \right] \\
    &= \sum_{x \in \mathcal{X}} P_{X}(x) D(P_{Y|X=x}||Q_{Y|X=x}) \: (disc.) \\
    &= \int P_{X}(x) D(P_{Y|X=x}||Q_{Y|X=x}) dx \: (cont.)
\end{align*}

Evidently, KL divergence is a special case of \textit{f}-divergence \cite{notes}. With $P_X \ll Q_X$, \textit{f}-divergence is defined by:
\begin{align*}
    D_{\textit{f}}(P_{X}||Q_{X}) &= \mathbb{E}_{Q_X} \left[ \textit{f} \left (\frac{dP_X}{dQ_X} \right ) \right] \\
    &= \sum_{x \in \mathcal{X}} Q_{X}(x) \: \textit{f} \left( \frac{P_{X}(x)}{Q_{X}(x)}\right) \: (discrete) \\
    &= \int Q_{X}(x) \: \textit{f} \left( \frac{P_X(x)}{Q_X(x)} \right ) dx \: (continuous)
\end{align*} Using the definition above, we can rewrite $D_{KL}(P_{X} || Q_{X})$ as:
\begin{align*}
    D_{\textit{f}}(P_{X}||Q_{X}) &= \mathbb{E}_{P_X} \left[ \log \frac{P_X}{Q_X} \right]  = \mathbb{E}_{Q_X} \left[ \frac{P_X}{Q_X} \log \frac{P_X}{Q_X} \right] \\
\end{align*} 
with $f\left(P_X / Q_x \right) = P_X / Q_X \log P_X / Q_X$. Another case of \textit{f}-divergence including Jensen-Shannon divergence ($JS(P_X || Q_X) = D_{KL}(P_X || (P_X + Q_X) / 2) + D_{KL}(Q_X || (P_X + Q_X) / 2)$), total variation ($T(P_X, Q_X) = 1/2 \: \mathbb{E}_{Q_X} \left[ |P_X / Q_X - 1| \right]$), etc \cite{notes}.  

\section{Mutual Information : Definitions and Properties}

\subsection{General Definition of Mutual Information}
We have discussed entropy in the previous section. We then define mutual information (MI) which quantifies the amount of information of a particular random variable given another random variable \cite{cover-2006, bishop}. Given joint probability $P_{X, Y}$ and marginal probability $P_{X}$ \& $P_{Y}$, mutual information between random variable $X$ and $Y$ is written by:
\begin{align}
   I(X; Y) &= \mathbb{E}_{P_{X, Y}} \log \frac{P_{X, Y}(x, y)}{P_{X}(x) P_{Y}(y)} \label{eq:MI-1} \\ 
   &= D(P_{Y|X} || P_{Y} | P_{X}) \label{eq:MI-2} \\ 
   &= D(P_{X|Y} || P_{X} | P_{Y}) \label{eq:MI-3} \\ 
   &= D_{KL}(P_{X, Y}(x, y) || P_{X}(x) P_{Y}(y)) \label{eq:MI-4} \\ 
   &= \sum_{x \in \mathcal{X}} \sum_{y \in \mathcal{Y}} P_{X, Y}(x, y) \log \frac{P_{X, Y}(x, y)}{P_{X}(x) P_{Y}(y)} \label{eq:MI-5} 
\end{align} Following the same constraint as entropy, MI can also be applied to a continuous random variable \cite{cover-2006}. In contrast to KL-divergence which is not symmetric, MI results in symmetric form means that $I(X; Y) = I(Y; X)$.

\begin{figure}
	\centering
	\includegraphics[width=0.5\textwidth, height=0.275\textheight]
		{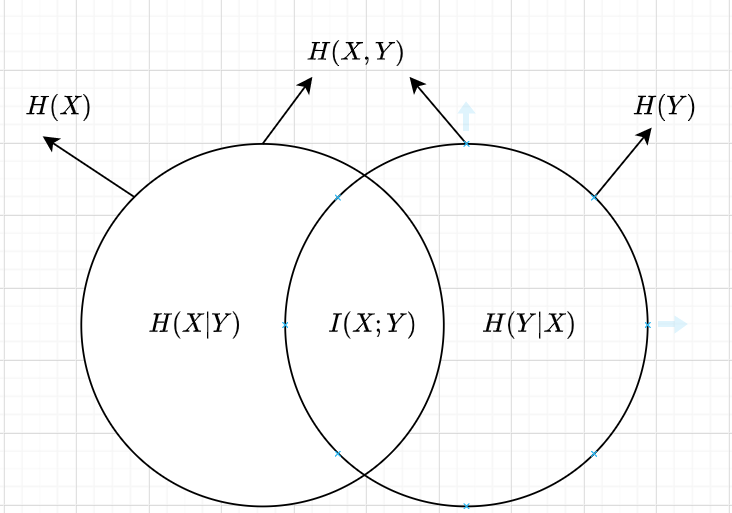}
	\caption{Venn diagram that shows the relationship between RV $X$ and $Y$. Observe that MI between $X$ and $Y$ is lied on the intersection between marginal entropy $H(X)$ and $H(Y)$ \cite{cover-2006}.}
	\label{fig:venn}
\end{figure}

In the previous section we already elaborate the entropy of joint distribution and conditional distribution as well. Evidently, those entropies have relationship with mutual information. \textcolor{red}{Figure \ref{fig:venn}} shows the relationship between two random variables from information theory perspective. From the figure, we can derive the definition of mutual information $I(X; Y)$ in term of $H(X)$, $H(Y)$, $H(X, Y)$, $H(Y|X)$, and $H(X|Y)$.

\begin{theorem}
\begin{align}
   I(X; Y) &= H(X) - H(X|Y) \label{eq:MI-entropy-1} \\
   &= H(Y) - H(Y|X) \label{MI-entropy-2} \\
   &= H(X) + H(Y) - H(X, Y) \label{eq:MI-entropy-3}
\end{align}
\end{theorem} \noindent Observe that $I(X; X) = H(X)$ for discrete RV (since $H(X, X) = H(X)$), otherwise it results $\infty$. We also can use the entropy to define the conditional mutual information. In particular, conditional MI of RV $X$ and $Y$ given $Z$ is defined by:
\begin{align}
   I(X; Y | Z) &= H(X|Z) - H(X|Y, Z) \label{eq:cond-MI-1} \\
   &= \mathbb{E}_{P_{X, Y, Z}} \log \frac{P_{X, Y | Z}(x, y | z)}{P_{X|Z}(x | z) P_{Y|Z}(y | z)} \label{eq:MI-entropy-2}
\end{align}. Mutual information also satisfied a chain rule theorem.
\begin{theorem}\label{MI-chain-rule-theorem}
\begin{align}
   I(X_1, ..., X_n; Y) = \sum I(X_i; Y | X_{i-1}, ..., X_1)
\end{align}
\end{theorem} 

We have discuss about the definition of MI in term of entropy and KL-divergence as well. In the next sections, we discuss about some properties of MI.  

\subsection{Convexity and Continuity of Mutual Information}
\begin{figure}
	\centering
	\includegraphics[width=0.5\textwidth, height=0.275\textheight]
		{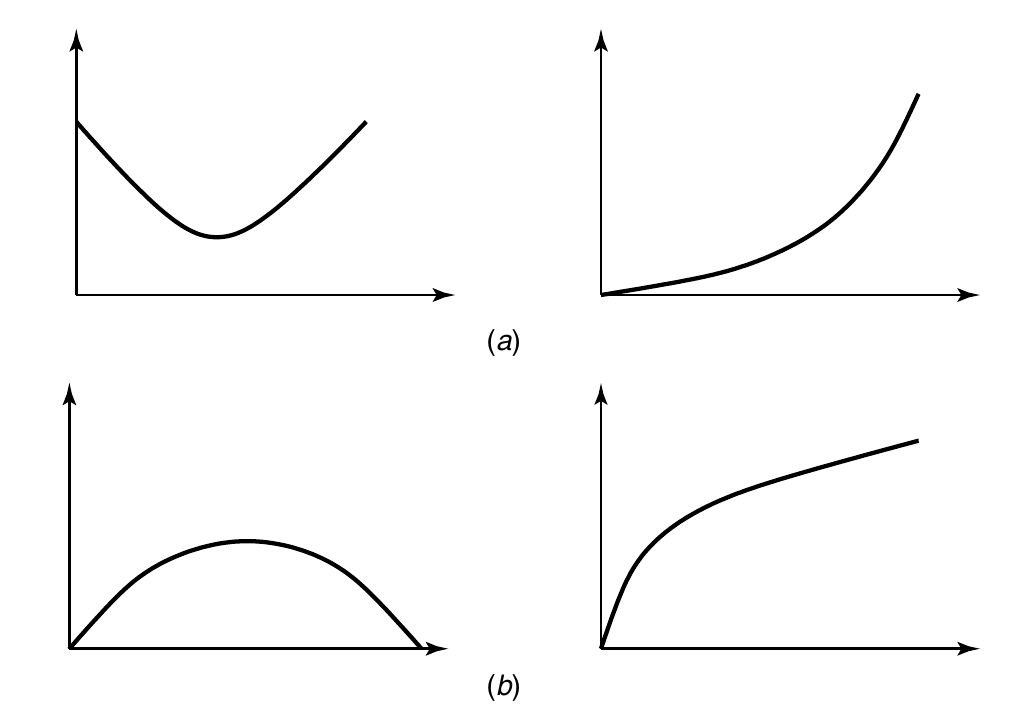}
	\caption{\textbf{a.} Convex functions represented by a an upward-opening. \textbf{b.} Concave function represented by a downward-opening curve. \cite{cover-2006}}
	\label{fig:convex-concave}
\end{figure}

We begin this section by defining convex and concave function. A function $f(x)$ is a convex function for interval $(u, v)$ if for every $x_i, x_j \in (u, v)$ and $0 \leq \alpha \leq 1$ holds $f(\alpha x_i + (1 - \alpha)x_j) \leq \alpha f(x_i) + (1 - \alpha) f(x_j)$ \cite{cover-2006}. We then call $f$ as strictly convex if equality is satisfied when $\alpha = 0$ or $\alpha = 1$. Meanwhile, a function $f$ is said to be concave when the negation $-f$ is convex. \textcolor{red}{Figure \ref{fig:convex-concave}} shows the examples of convex and concave function.

 \noindent

We then have three theorems about the convexity and concavity of KL-divergence, entropy, and mutual information.

\begin{theorem}\label{KL-convex-theorem}
    $D_{KL}(P_X || Q_X)$ is convex function. In particular given the pair of distribution functions $(P_Xi, Q_Xi)$ and $(P_Xj, Q_Xj)$ then
\begin{align}
    D_{KL}(\alpha P_{Xi} + (1 - \alpha) P_{Xj} || \alpha Q_{Xi} + (1 - \alpha)Q_{Xj}) \leq \notag \\ \alpha D_{KL}(P_{Xi} || Q_{Xi} + (1 - \alpha)D_{KL}(P_{Xj}||Q_{Xj}))  \label{eq:KL-convex}    
\end{align} \noindent for $0 \leq \alpha \leq 1$ \cite{cover-2006}
\end{theorem}

\begin{theorem}\label{entropy-concave-theorem}
    Given a probability distribution $P_X$ of RV $X$ on space $\mathcal{X}$, entropy $H(P_X)$ is concave \cite{cover-2006}. 
\end{theorem}

\begin{theorem}\label{MI-convex-concave-theorem}
    Let $(X, Y) \sim P_{X, Y}(x, y) = P_{X}(x)P_{Y|X}(y | x)$. The mutual information $I(X; Y)$ is a concave function of $P_{X}(x)$ for fixed $P_{Y|X}(y | x)$ and a convex function of $P_{Y|X}(y | x)$ for fixed $P_{X}$ \cite{cover-2006}.  
\end{theorem}

Besides being convex, MI also possesses continuity property. We show this property by first seeing that KL divergence and entropy are continuous. Formally, for a fix distribution $Q_X$ on space $\mathcal{X}$ with $Q(x) > 0 \quad \forall x \in \mathcal{X}$ then $D_{KL}(P_X || Q_X)$ is continuous. In particular, $H(P_X)$ is continuous \cite{notes}. We then define MI by $I(X; Y) = H(X) + H(Y) - H(X, Y)$. Since $H(X)$ is continuous, then $I(X; Y)$ is assured to be continuous. 

\subsection{Jensen Inequality and The Consequences for Mutual Information}
 The Jensen inequality requires a function to be convex. 
\begin{theorem}\label{jensen-inequality-theorem}
Jensen's inequality: if f is a convex function and $X$ is a random variable then
\begin{align}
\mathbb{E}(g(X)) \geq g(\mathbb{E} X)
\end{align}
with equality hold when the function is strictly convex
\end{theorem} \noindent This inequality is used to discover the property of KL-divergence. Note that we have shown that KL-divergence is a convex function (\textcolor{blue}{Equation \ref{eq:KL-convex}}).
\begin{theorem}
Divergence inequality: Given distribution function $P_{X}$ and $Q_{X}$ over $x \in \mathcal{X}$. Then it applies that
\begin{align}
D(P_{X} || Q_{X}) \geq 0
\end{align}
with equality hold when $P_{X}(x) = Q_{X}(x)$
\end{theorem} 

We then use the theorem above to imply the property of MI. Since $I(X; Y) = D(P_{X, Y}(x, y) || P_{X}(x)P_{Y}(y))$ then it implies that $I(X; Y) \geq 0$ with equality hold when $P_{X, Y}(x, y) = P_{X}(x) P_{Y}(y)$. Second implication is $I(X; Y | Z) \geq 0$ since we can transform it into the form of $D_{KL}$ as well. The last implication already being stated in preliminary section which is $H(X|Y) \leq H(X)$. Recall that $I(X; Y) = H(X) - H(X|Y)$. Since $I(X; Y) \geq 0$ then $H(X) - H(X|Y) \geq 0$.

\subsection{Relations between Conditional Independence and Mutual Information}
In this section, we show that some conditional independent forms of distribution results in inequality of MI. Random variable $X, Y, Z$ are said to be conditional independent if:
\begin{align}
    P_{X, Z | Y}(x, z | y) &= \frac{P_{X, Y, Z}(x, y, z)}{P_{Y}(y)} \notag \\ 
    &= \frac{P_{X, Y}(x, y) P_{Z|Y}(z|y)}{P_Y(y)} \notag \\ 
    &= P_{X|Y}(x|y)P_{Z|Y}(z|y)
\end{align} From the graphical model perspective, random variable $X, Z$ are conditionally independent given $Y$ if and only if $X, Y, Z$ forms a Markov chain denoted by $X \rightarrow Y \rightarrow Z$ \cite{notes}. Under the circumstance, joint probability $X, Y, Z$ is defined by:
\begin{align}
    P_{X, Y, Z} = P_{X}(x)P_{Y|X}(y | x)P_{Z|Y}(z|y)
\end{align} \noindent Furthermore, Markov chain $X \rightarrow Y \rightarrow Z$ also implies $Z \rightarrow Y \rightarrow X$ \cite{notes}. Another form of Markov chain that satisfies conditional independence is $X \leftarrow Y \rightarrow Z$ \cite{notes} where the joint probability is defined by:
\begin{align}
    P_{X, Y, Z} = P_{Y}(y)P_{X|Y}(x | y)P_{Z|Y}(z|y)
\end{align} \noindent Having the definitions, we derive inequality theorem constrained by the Markov chain form.

\begin{theorem}\label{markov-theorem}
    if $X \rightarrow Y \rightarrow Z$ then $I(X; Y) \geq I(X; Z)$
\end{theorem}

\noindent Using the above theorem, we can derive two properties. First, if $Z = g(Y)$ then we have $I(X; Y) \geq I(X; g(Y))$ since $X \rightarrow Y \rightarrow g(Y)$ will follows Markov chain. We also have $(X; Y | Z) \leq I(X; Y)$. This property comes by noticing that $I(X; Y | Z) = 0$ and $I(X; Z) \geq 0$ \cite{notes}.

\subsection{Geometric Interpretation of Mutual Information}
We know elaborate mutual information from the perspective of geometry. First, we examine mutual information as conditional divergence. Recall \textcolor{blue}{Equation \ref{eq:MI-2}}, we write it into discrete form as:
\begin{align*}
    I(X; Y) &= D_{KL}(P_{Y|X} || P_{Y} | P_{X}) \\
            &= \sum_{x} D_{KL}(P_{Y|X = x} || P_{Y}) P_X(x)    
\end{align*} \noindent We can see that each outcome $x$ is weighted by probability distribution $P_X(x)$. Hence, we can say that MI is a weighted distance measure between two distributions.

In this section, we specify an auxiliary distribution $Q$ to redefine MI.
\begin{theorem}\label{golden-theorem}
$\forall Q_{Y}$ such that $D_{KL}(P_Y || Q_Y) < \infty$
\begin{align}
    I(X; Y) = D_{KL}(P_{X|Y} || Q_{X} | P_{Y}) - D(P_{X} || Q_{X}) \label{eq:MI-geometric-1} 
\end{align}
\end{theorem}  \noindent If $Q_{X}$ is optimum such that $Q_{X} = P_{X}$ then the second term can be removed, thus $I(X; Y) = \underset{Q_X}{\argmin}\, D_{KL}(P_{X|Y} || Q_{X} | P_{Y})$ \cite{notes}. Intuitively, the auxiliary distribution $Q_{X}$ will be moving towards the real distribution $P_X$ in some probability measure space during the optimization. 

We can scale up the utilization of auxiliary/variational distribution for two RV $X, Y$. In the theorem below, we specify a new auxiliary distribution $Q_Y$.
\begin{theorem}\label{distance-product-theorem}
    We can see mutual information as a distance to product distribution \cite{notes}.
    \begin{align}
        I(X; Y) = \argmin_{Q_X, Q_Y} D_{KL}(P_{X, Y} || Q_X Q_Y) \label{eq:MI-geometric-2}
    \end{align}
\end{theorem} \noindent We can generalize the theorem above to conditional mutual information as  $I(X; Z | Y) = \underset{{Q_{X, Y ,Z}:X \rightarrow Y \rightarrow Z}}{\argmin} D_{KL}(P_{X, Y, Z} || Q_{X, Y, Z})$ \cite{notes}.

\subsection{Variational Form of Mutual Information}
In the previous section, we have discussed one of the variational form of MI (\textcolor{blue}{Equation \ref{eq:MI-geometric-1}}). This section provides another two variational forms of MI. These forms are based on characterizations KL-divergence : Donsker-Varadhan and Gelfand-Yaglom-Perez.

We begin by introducing the Donsker-Varadhan form of KL-divergence.
\begin{theorem}\label{donsker-varadhan-theorem}
    Donsker-Varadhan: Let $P_X, Q_X$ be a probability measures of RV $X$ on space $\mathcal{X}$ and $\mathcal{C}$ be the set of function $g : \mathcal{X} \rightarrow \mathbb{R}$ such that $\mathbb{E}_{Q_X}[e^{g(X)}] < \infty$. If $D_{KL}(P_X || Q_X) < \infty$ then for all $f \in \mathcal{C}$ expectation $\mathbb{E}_{P_X}[g(X)]$ exists and also \cite{notes}:
    \begin{align}
        D(P_X || Q_X) = \underset{g \in \mathcal{C}}{\sup} \, \mathbb{E}_{PX}[g(X)] - \log \mathbb{E}_{Q_X}[e^{g(X)}] \label{eq:KL-DKV}
    \end{align}
\end{theorem} \noindent We then apply the theorem above to find the Donsker-Varadhan form of MI. Using the \textcolor{blue}{Equation \ref{eq:MI-4}} and \textcolor{blue}{Equation \ref{eq:KL-DKV}}  we get:

\begin{align}
    I(X; Y) = \underset{g}{\sup} \, \mathbb{E}[g(X, Y)] - \log \mathbb{E}[e^{g(X, \hat{Y})}]
\end{align} \noindent with $\hat{Y}$ is a duplicate of $Y$ which is independent of $X$ and the supremum is over bounded or even bounded by continuous functions $g$. 

The next theorem introducing Gelfand-Yaglom-Perez form of KL-divergence which involves $\sigma$-space.
\begin{theorem}
    Gelfand-Yaglom-Perez: Let $P_X, Q_X$ be a probability measures on space $\mathcal{X}$ with $\sigma$-algebra $\mathcal{F}$. Then:
    \begin{align}
        D(P_X || Q_X) = \underset{\{E_1, ..., E_n \}}{\sup} \sum_{i = 1}^{n} {P_X[E_i]} \log \frac{P_X[E_i]}{Q_X[E_i]} \label{eq:KL-GYP}
    \end{align} with the supremum is over all finite $\mathcal{F}$-measurable partitions: $\cup_{j = 1}^{n} E_j = \mathcal{X}$, $E_j \cap E_i = \emptyset$.
\end{theorem} \noindent with $0 \log \frac{1}{0} = 0$ and $\log \frac{1}{0} = \infty$ for conventions. We then apply the theorem above to find the Donsker-Varadhan form of MI. Using the \textcolor{blue}{Equation \ref{eq:MI-4}} and \textcolor{blue}{Equation \ref{eq:KL-GYP}}  we get:

\begin{align}
    I(X; Y) = \underset{\{E_i\} \times \{F_j\}}{\sup} \, \sum_{i, j} P_{X, Y} \left[ E_i \times F_j \right] \log \frac{P_{X, Y}\left[ E_i \times F_j\right]}{P_X\left[ E_i \right] P_Y\left[ F_j\right]}
\end{align} \noindent with supremum is over finite partitions space $\mathcal{X}$ and $\mathcal{Y}$.

\section{Mutual Information: Estimation Methods}
We already know that mutual information can capture the dependence of random variables. But often times we can not directly use the closed function of mutual information. Recall that in the \textcolor{blue}{Equation \ref{eq:MI-1}}, we need the access to $P_{X, Y}(x, y)$, $P_{X}$, and $P_Y(y)$ which are not always guaranteed. The mutual information estimation then come to bound the true MI. The estimation is either upper-bounding or lower-bounding the true MI. The idea of MI estimations come from variational form of MI. In the previous section we already discuss three variational forms of MI. We try to approximate the MI estimation by using an auxilary distribution or a critic function. 

\begin{figure}
	\centering
	\includegraphics[width=0.5\textwidth, height=0.275\textheight]
		{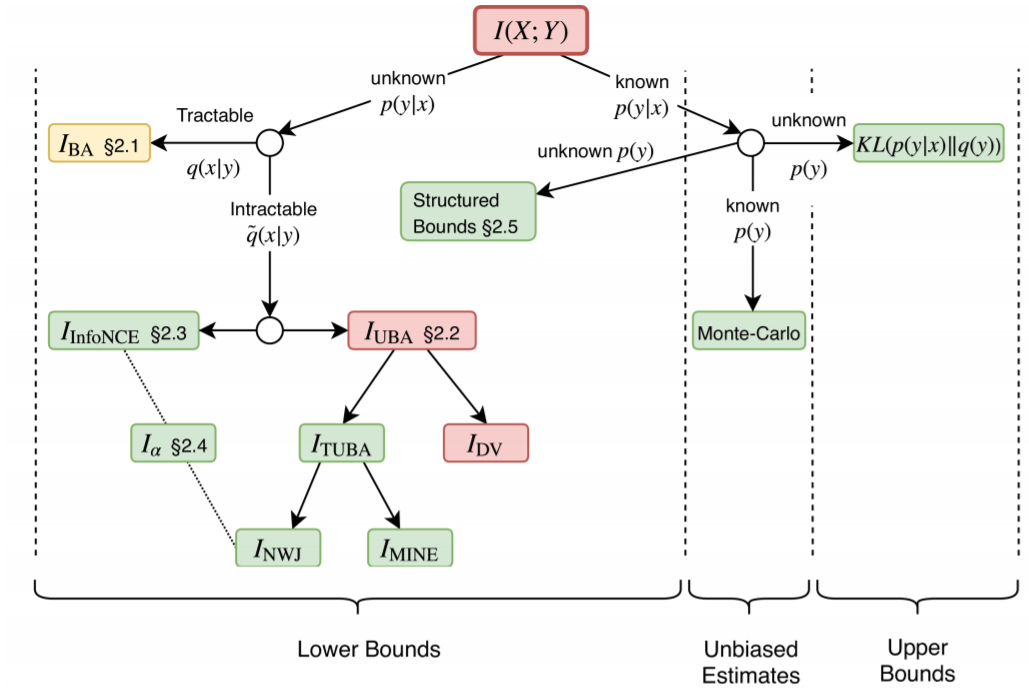}
	\caption{Schematic of variational bounds of mutual information proposed by Poole et al., 2019. The schematic is based on the presence of the available distributions \cite{on-variational-bounds}}
	\label{fig:variational-bounds}
\end{figure}

In this section, we review several MI estimation methods. The review has been conducted before by Poole et al., (2019). \textcolor{red}{Figure \ref{fig:variational-bounds}} shows the schematic of variational bounds of mutual information proposed by Poole et al., 2019 \cite{on-variational-bounds}. In this article, we divide the reviews into three sections: normalized bounds, unnormalized bounds, and improved bounds.

\subsection{Normalized Bounds}
In this section, we discuss two versions of normalized bounds, upper bound and lower bound MI estimation. The bounds were firstly introduced by Agakov \cite{barber-agakov}. Recall the definition of MI in \textcolor{blue}{Equation \ref{eq:MI-1}}. We then rewrite $P_{X, Y}(x, y) = P_{Y | X}(y | x) P_Y(y)$. Subsequently, we apply \textcolor{orange}{Theorem \ref{distance-product-theorem}} by replacing $P_{Y}(y)$ with a variational distribution $Q_{Y}(y)$. Mathematically, we can write:

\noindent \begin{align}\label{eq:ba-upper}
    I(X, Y) &= \mathbb{E}_{P_{X, Y}(x, y)} \left [ \log \frac{P_{Y | X}(y | x)}{P_{Y}(y)} \right ]  \notag \\
            &= \mathbb{E}_{P_{X, Y}(x, y)} \left[ \log \frac{P_{Y | X}(y | x) Q_{Y}(y)}{P_{Y}(y) Q_{Y}(y)} \right ] \notag \\
            &= \mathbb{E}_{P_{X, Y}(x, y)} \left[ \log \frac{P_{Y | X}(y | x)}{Q_{Y}(y)} \right] - KL(P_{Y}(y) || Q_{Y}(y)) \notag \\
            &\geq \mathbb{E}_{P_{X, Y}(x, y)} \left[ \log \frac{P_{Y | X}(y | x)}{Q_{Y}(y)} \right] \triangleq I_{R}
\end{align} \noindent Thus, we upper-bounding the MI. Note that in \textcolor{blue}{Theorem \ref{distance-product-theorem}}, we can assure equality since we assumed that we can find the optimum $Q_{Y}$. We also need to constraint $Q_{Y}(y)$ to be intractable. However, the assumption is not assured in the real world. One of the applications of the bound is for deep information bottle-neck model \cite{information-bottleneck}.

In contrast, we derive lower-bound by applying \textcolor{orange}{Theorem \ref{distance-product-theorem}} into the numerator $P_{X | Y}(x | y)$ \cite{barber-agakov}. We replace $P_{X | Y}(x | y)$ with $Q_{X | Y}(x | y)$:
\noindent \begin{align}\label{eq:ba-lower}
    I(X, Y)  &= \mathbb{E}_{P_{X, Y}(x, y)} \left [ \log \frac{P_{X | Y}(x | y)}{P_{X}(x)} \right ] \notag \\
             &= \mathbb{E}_{P_{X, Y}(x, y)} \left [ \log \frac{Q_{X | Y}(x | y)}{P_{X}(x)} \right ] + \notag \\
             & \quad \: \mathbb{E}_{P_{Y}(y)}\left[ KL(P_{X | Y}(x|y) || Q_{X | Y}(x|y)) \right]  \notag \\
             &\geq \mathbb{E}_{P_{X, Y}(x, y)} \left [ \log Q_{X | Y}(x | y) \right ] + h(X) \triangleq I_{BA}
\end{align} \noindent with $h(X)$ is the marginal entropy of $X$. The objective is tractable if $h(X)$ is known. However, $h(X)$ is often to be unknown. This bound has been applied as regularizer of Info-GAN objective function \cite{infogan}.

\subsection{Unnormalized Bounds}
We can solve the intractibility problem from the previous section by using the unnormalized form of $Q_{X | Y}(x | y)$. We write the distribution in terms of a critic function $g(x, y)$ and marginal distribution $P_X(x)$:

\noindent \begin{align}\label{eq:unnormalized}
    Q_{X|Y}(x | y) = \frac{P_X(x)}{Z(y)} e^{g(x, y)}; \:Z(y) = \mathbb{E}_{P_{X}(x)} \left[ e^{g(x, y)} \right]
\end{align}

\noindent By applying the equation above into \textcolor{blue}{Equation \ref{eq:ba-lower}}, we get unnormalized BA estimation ($I_{UBA}$): 
\noindent \begin{align}\label{eq:UBA}
    \mathbb{E}_{P_{X, Y}(x, y)}[g(x, y)] - \mathbb{E}_{P_Y(y)}\left[ \log Z(y) \right] \triangleq I_{UBA}
\end{align} \noindent Note that in the equation above, the entropy $H(X)$ is no longer involved. However, the term $\log Z(y)$ is still intractable. Since log function is convex, by applying the Jensen inequality we have Donsker-Varadhan lower bound \cite{DKV}:

\noindent \begin{align}\label{eq:DKV-lower-bound}
    \mathbb{E}_{P_{X, Y}(x, y)}[g(x, y)] - \log \mathbb{E}_{P_Y(y)}\left[ Z(y) \right] \triangleq I_{DKV}
\end{align} \noindent Note that $I_{BA} \geq I_{DKV}$ (by Jensen inequality). We have seen this form from \textcolor{orange}{Theorem \ref{donsker-varadhan-theorem}}, except without confirming the equality. This bound is also still intractable. By upper-bounding the log partition $\log Z(y)$, we can form a tractable bound. We specify an inequality $\log(x) \leq \frac{x}{a} + \log(a) - 1, \forall x, a > 0$. Applying the inequality into the second term of \textcolor{blue}{Equation \ref{eq:UBA}} will give $\log (Z(y)) \leq \frac{Z(y)}{a(y)} + \log(a(y)) - 1$. Finally, we can rewrite the bound as:
\noindent \begin{align}\label{eq:TUBA}
    &\quad \mathbb{E}_{P_{X, Y}(x, y)}[g(x, y)] - \notag \\
    &\quad \mathbb{E}_{P_Y(y)}\left[ \frac{\mathbb{E}_{P_X(x)}\left[ e^{g(x, y)} \right]}{a(y)} + \log(a(y)) - 1 \right] \triangleq I_{TUBA}
\end{align} \noindent The bound is optimized with respect to $a(y)$ and $g$. Both are optimized simultaneously. Furthermore, we can simplify \textcolor{blue}{Equation \ref{eq:TUBA}} by set $a(y) = e$ which leads to Nguyen-WainWright-Jordan estimation \cite{NWJ}:  

\noindent \begin{align}\label{eq:NWJ}
    \mathbb{E}_{P_{X, Y}(x, y)}[g(x, y)] -  e^{-1} \mathbb{E}_{P_Y(y)}\left[ Z(y) \right] \triangleq I_{NWJ}
\end{align} \noindent Generally, unnormalized bounds suffer from the high variance problem due to the log partition function.

\subsection{Improved Bounds}
In this section, we discuss several improvements that have been made to respond the current drawbacks of normalized and unnormalized bound. 

Info-NCE extends the NWJ estimations by using Monte Carlo estimation on multiple samples \cite{cpc}:

\noindent \begin{align}\label{eq:infonce}
    I(X, Y) \geq \mathbb{E}\left [ \frac{1}{k} \sum_{i=1}^{K} \log \frac{e^{f(x_{i}, y_{i})}}{\sum_{j=1}^{K} e^{f(x_j, y_j)}} \right ] \triangleq I_{NCE}
\end{align} \noindent However, this estimation tends to have a higher bias compared to NWJ estimation. 

Barber-Agakov upper bound estimation also have a problem with the variational distribution $Q_Y(y)$. Evidently, learning distribution $Q_{y}(y)$ without any prior knowledge is extremely difficult especially when RV $Y$ is high dimensional \cite{club, problem-ba-upper}. The distribution $Q_{Y}(y_i)$ can be replaced with Monte Carlo approximation $Q_{y}(y) = \frac{1}{K - 1} \underset{j \neq i}{\sum}P_{Y | X}(y | x_j)$ \cite{on-variational-bounds}, we derive one left out (L1-out) upper bound estimation:

\noindent \begin{align}\label{eq:infonce}
    \mathbb{E} \left[ \frac{1}{K} \sum_{i = 1}^{K} \left[ \log \frac{P_{Y | X}(y_i | x_i)}{\frac{1}{K - 1} \sum_{j \neq i} P_{Y|X}(y_i | x_j)} \right] \right] \triangleq I_{L1-out}
\end{align} \noindent The estimation method is called one left out because we discard one sample on the denumerator inside the sum. The drawback of this method lies to its numerical instability especially when RV $Y$ is high dimensional \cite{club}.

Given all existing MI estimations, the current methods still have several drawbacks. MI estimation is currently active research. For example, current research shows that we can estimate MI by using optimal transport concept that is Wasserstein distance \cite{wasserstein-mi}. Another research using clipping method to reduce the variance of NWJ estimation \cite{smile}.

\section{Conclusion}
The article discussed the definitions of mutual information in the form of KL-divergence and entropy as well. The article then delivered some properties of mutual information including concavity, the continuity, Jensen inequality, conditional independence, and variational form. Later, the article reviewed several mutual information estimation methods. The estimation methods are useful whenever we have an unaccessible probability (commonly marginal distribution). We also mention that the current mutual information estimation methods also have drawbacks.

\appendix
\section{Proof of Theorem on Entropy Section}

\subsection{Proof of Theorem \ref{entropy-chain-rule}}

We prove the theorem in the discrete form of random variables \cite{cover-2006}
\begin{align*}
    H(X, Y) &= - \underset{x \in \mathcal{X}}{\sum} \: \underset{y \in \mathcal{Y}}{\sum} P_{X, Y}(x, y) \log P_{X, Y}(x, y) \\
    &= - \underset{x \in \mathcal{X}}{\sum} \: \underset{y \in \mathcal{Y}}{\sum} P_{X, Y}(x, y) \log P_{X}(x) P_{Y|X}(y |x) \\
    &= - \underset{x \in \mathcal{X}}{\sum} \: \underset{y \in \mathcal{Y}}{\sum} P_{X, Y}(x, y) \log P_{X}(x) \\
    & \quad - \underset{x \in \mathcal{X}}{\sum} \: \underset{y \in \mathcal{Y}}{\sum} P_{X, Y}(x, y) \log P_{Y|X}(y |x) \\
    &= - \underset{x \in \mathcal{X}}{\sum} P_{X} \log P_X(x) \\
    & \quad - \underset{x \in \mathcal{X}}{\sum} \: \underset{y \in \mathcal{Y}}{\sum} P_{X, Y}(x, y) \log P_{Y|X}(y |x) \\
    &= H(X) + H(Y|X)
\end{align*}

\section{Proof of Theorem on General Definition of Mutual Information Section}

\subsection{Proof of Theorem \ref{entropy-chain-rule}}
\begin{align*}
    I(X_1, ..., X_n; Y) &= H(X_1, ..., X_n) - H(X_1, ..., X_n | Y) \\
    &= \sum_{i=1}^{n} H(X_i | X_{i - 1}), ..., X_1) \\ 
    & \quad - \sum_{i = 1}^{n} H(X_i | X_{i - 1}, ..., X1, Y) \\
    &= \sum_{i = 1}^{n} I(X_i; Y | X1, ..., X_{i - 1})
\end{align*}

\section{Proof of Theorems on Convexity and Continuity of MI Section}

\subsection{Proof of Theorem \ref{KL-convex-theorem}}
In order to prove the theorem, we apply log sum inequality on the left-hand side \cite{cover-2006}.
\begin{align*}
    &(\alpha P_1(x) + (1 - \alpha)P_2(x) \log \frac{\alpha P_1(x) + (1 - \alpha)P_2(x)}{\alpha Q_1(x) + (1 - \alpha)Q_2(x)}) \\
    &\leq \alpha P_1(x) \log \frac{\alpha P_1(x)}{\alpha Q_1(x)} + (1 - \alpha) P_2(x) \log \frac{(1 - \alpha) P_2(x)}{(1 - \alpha) Q_2(x)}
\end{align*}

\subsection{Proof of Theorem \ref{entropy-concave-theorem}}
The result comes from the fact that $H(P_X) = log |\mathcal{X}| - D_{KL}(P_X || U_X)$ where $U_X$ is an uniform distribution of $x \in \mathcal{X}$. The negative term of KL-divergence of the equation then implies its concavity \cite{cover-2006}. 

\subsection{Proof of Theorem \ref{MI-convex-concave-theorem}}
We recall the definition of MI to prove the theorem:
\begin{align*}
    I(X; Y) &= H(Y) - H(Y | X) \\
    &= H(Y) - \sum_{x} P_{X}(x) H(Y | X = x)
\end{align*}

\noindent First, we proof the first argument of the theorem. Given $P_{Y|X}(y | x)$, then $P_{Y}(y)$ is linear function of $P_{X}(x)$. Since $H(Y)$ is a convex function of $P_Y(y)$, then we can say that $H(Y)$ is a concave function of $P_X(x)$. We can see the second term of as a function of $P_X(x)$. Thus, the difference is a concave function of $P_{X}(x)$ \cite{cover-2006}.

For the second argument, we specify two conditional distributions $P_{1\:Y|X}, P_{2 \: Y|X}$. The corresponding joint distributions given the conditional distributions are $P_{1 \: X, Y}(x, y) = P_X(x)P_{1 \: Y|X}(y | x)$ and $P_{2 \: X, Y}(x, y) = P_X(x)P_{2\:Y|X}(y | x)$ with respective marginals $P_{X}(x)$, $P_{1 \: Y}(y)$ and $P_{X}(x)$, $P_{1 \: Y}(y)$ We then specify a conditional distribution which is a mixture of $P_{1 \: Y|X}(y | x)$ and $P_{2 \: Y|X}(y | x)$:
\begin{align*}
    P_{\alpha Y|X}(y | x) = \alpha P_{1 \: Y|X}(y | x) + (1 - \alpha) P_{2\:Y|X}(y | x)
\end{align*} \noindent $0 \leq \alpha \leq 1$. We can easily see that the corresponding joint distribution is also a mixture joint distribution,
\begin{align*}
    P_{\alpha X, Y}(x, y) = \alpha P_{1 \: X, Y}(x, y) + (1 - \alpha) P_{2\:X, Y}(x, y)
\end{align*} \noindent and the marginal distribution $Y$ is also a mixture,
\begin{align*}
    P_{\alpha Y}(y) = \alpha P_{1 \:Y}(y) + (1 - \alpha) P_{2\: Y}(y)
\end{align*} If we let $Q_{\alpha \: X, Y}(x, y) = P_{X}(x)P_{\alpha \: Y}(y)$ be the product of the marginal distributions, then we have:
\begin{align*}
    Q_{\alpha X, Y}(x, y) = \alpha Q_{1 \: X, Y}(x, y) + (1 - \alpha) Q_{2\:X, Y}(x, y)
\end{align*} \noindent We already know that MI can be thought as KL-divergence between joint distribution and the product of marginal distributions, hence:
\begin{align*}
    I(X; Y) = D_{KL}(P_{\alpha \: X, Y}(x, y) || Q_{\alpha |: X, Y}(x, y))
\end{align*} Since KL-divergence is a convex function, thus the MI is convex function of conditional distribution \cite{cover-2006}.

\section{Proof of Theorems on Jensen Inequality and The Consequences for MI}

\subsection{Proof of Theorem \ref{jensen-inequality-theorem}}
The proof is for discrete distribution by using induction on the number of mass point. At first, we settle the base case which is the inequality of two-mass distribution ($x_1$ and $x_2$) \cite{cover-2006}. Let $w_1$ and $w_2$ be the weights for $x_1$ and $x_2$ respectively, the inequality becomes:
\begin{align*}
    w_1\,g(x_1) + w_2\,g(x_2) \geq g(w_1\,x_1 + w_2\,x_2)
\end{align*} Note that this inequality is similar with the definition of convex function. Suppose that the inequality is true for $k - 1$ points. If we write $\hat{w_i} = w_i / (1 - w_k)$ then :
\begin{align*}
    \sum_{i = 1}^{k}w_i\,g(x_i) &= w_k\,g(x_k) + (1 - w_k) \sum_{i = 1}^{k - 1}\hat{w_i}\,g(x_i) \\
    &\geq w_k g(x_k) + (1 - w_k)g\left( \sum_{i = 1}^{k - 1} \hat{w_i}x_i\right) \\
    &\geq g\left( w_k\,x_k + (1 - w_k)\sum_{i = 1}^{k - 1}\hat{w_i}x_i\right) \\
    &= g \left( \sum_{i = 1}^{k} w_i\,x_i\right)
\end{align*}

\subsection{Proof of Theorem \ref{KL-convex-theorem}}
Let $\mathcal{A}$ be the support of $P_{x}(x)$
\begin{align*}
    - D_{KL}(P_X || Q_X) &= - \sum_{x \in \mathcal{A}} P_X(x) \log \frac{P_X(x)}{Q_X{x}} \\
    &= \sum_{x \in \mathcal{A}} P_X(x) \log \frac{Q_X(x)}{P_X(x)} \\
    &\leq \log \sum_{x in \mathcal{A}} P_{X}(x) \log \frac{Q_X(x)}{P_{X}(x)} \\
    &= \log \sum_{x \in \mathcal{A}} Q_X(x) \\
    &\leq \log \sum_{x \in \mathcal{X}} Q_{X}(x) \\
    &= \log 1 \\
    &= 0
\end{align*}

\section{Proof of Theorem on Section Relations between Conditional Independence and MI}

\subsection{Proof of Theorem \ref{markov-theorem}}
\begin{align*}
    I(X; Y, Z) &= I(X; Z) + I(X; Y | Z) \\
    &= I(X; Y) + I(X; Z | Y)
\end{align*} \noindent Since $X$ and $Z$ are conditionally independent given $Y$, thus $I(X; Z | Y)$. Moreover, $I(X; Y | Z) \geq 0$ implies:
\begin{align*}
    I(X; Y) \geq I(X; Z)
\end{align*}

\section{Proof of Theorems on Section Geometric Interpretation of MI}

\subsection{Proof of Theorem \ref{golden-theorem}}
\begin{align*}
    I(X; Y) &= \mathbb{E}_{P_{X, Y}} \left[ \log \frac{P_{Y|X}(y | x)}{P_{Y}(y)} \right] \\
    &= \mathbb{E}_{P_{X, Y}} \left[ \log \frac{P_{Y|X}(y | x) Q_Y(y)}{P_{Y}(y)Q_Y(y)} \right] \\
    &= \mathbb{E}_{P_{X, Y}} \left[ \log \frac{P_{Y|X}(y | x)}{Q_Y(y)} \right] + \\
    &\quad \mathbb{E}_{P_{X, Y}} \left[ \log \frac{Q_Y(y)}{P_Y(y)} \right] \\
    &= \mathbb{E}_{P_X} \mathbb{E}_{P_Y} \left[ \log \frac{P_{Y|X}(y | x)}{Q_Y(y)} \right] - \\
    &\quad \mathbb{E}_{P_{X, Y}} \left[ \log \frac{P_Y(y)}{Q_Y(y)} \right] \\
    &= \mathbb{E}_{P_X}\left[ D_{KL}(P_{Y|X} || Q_Y)\right] - D_{KL}(P_Y || Q_Y) \\
    &=  D_{KL}(P_{Y|X} || Q_Y | P_X) - D_{KL}(P_Y || Q_Y)
\end{align*}

\subsection{Proof of Theorem \ref{distance-product-theorem}}
Since $Q_X$ and $Q_Y$ minimum, we have $Q_X = P_X$ and $Q_Y = P_Y$.

\begin{align*}
    &I(X; Y) = \mathbb{E}_{P_{X, Y}} \left[ \log \frac{P_{X, Y}(x, y) Q_X(x) Q_Y(y)}{P_{X}(x) P_Y(y) Q_X(x) Q_Y(y)} \right] \\
    &= \mathbb{E}_{P_{X, Y}} \left[ \log \frac{P_{X, Y}(x, y)}{P_{X}(x) P_Y(y)} \right] + \mathbb{E}_{P_{X, Y}} \left[ \log \frac{Q_{X}(x)}{P_{X}(x)} \right] + \\
    &\quad  \mathbb{E}_{P_{X, Y}} \left[ \log \frac{Q_{Y}(y)}{P_Y(y)} \right] \\
    &= D_{KL}(P_{X, Y} || Q_X Q_Y) + 0 + 0 \\
    &= D_{KL}(P_{X, Y} || Q_X Q_Y)
\end{align*}

\end{document}